\documentclass[runningheads]{llncs}
\usepackage{epsfig}
\usepackage{graphicx}
\usepackage{amsmath}
\usepackage{amssymb}
\usepackage{pdfrender}
\usepackage{marvosym}
\usepackage{subfigure}
\usepackage{multirow}
\usepackage{array}
\usepackage{color}
\usepackage{tabu}
\usepackage{longtable}

\usepackage[breaklinks=true,letterpaper=true,colorlinks,bookmarks=false]{hyperref}
\newcommand{\etal}{\textit{et al}. }
\newcommand{\ie}{\textit{i}.\textit{e}., }
\newcommand{\eg}{\textit{e}.\textit{g}., }
\usepackage[T1]{fontenc}
\usepackage{graphicx}
\begin{document}
\title{Accurate and Robust Lesion RECIST Diameter Prediction and Segmentation with Transformers}
\titlerunning{Lesion RECIST Diameter Prediction and Segmentation with Transformers}

\author{Youbao Tang\inst{1}\textsuperscript{\Letter} \and
Ning Zhang\inst{1} \and
Yirui Wang\inst{1} \and
Shenghua He\inst{1} \and
Mei Han\inst{1} \and
Jing Xiao\inst{2} \and
Ruei-Sung Lin\inst{1}}


\authorrunning{Youbao Tang, \etal}

\institute{PAII Inc., Palo Alto, CA, USA \\ \email{tybxiaobao@gmail.com} \and Ping An Technology, Shenzhen, China}

\maketitle              
\begin{abstract}
Automatically measuring lesion/tumor size with RECIST (Response Evaluation Criteria In Solid Tumors) diameters and segmentation is important for computer-aided diagnosis. Although it has been studied in recent years, there is still space to improve its accuracy and robustness, such as (1) enhancing features by incorporating rich contextual information while keeping a high spatial resolution and (2) involving new tasks and losses for joint optimization. To reach this goal, this paper proposes a transformer-based network (MeaFormer, \textbf{Mea}surement trans\textbf{Former}) for lesion RECIST diameter prediction and segmentation (LRDPS). It is formulated as three correlative and complementary tasks: lesion segmentation, heatmap prediction, and keypoint regression. To the best of our knowledge, it is the first time to use keypoint regression for RECIST diameter prediction. MeaFormer can enhance high-resolution features by employing transformers to capture their long-range dependencies. Two consistency losses are introduced to explicitly build relationships among these tasks for better optimization. Experiments show that MeaFormer achieves the state-of-the-art performance of LRDPS on the large-scale DeepLesion dataset and produces promising results of two downstream clinic-relevant tasks, \ie 3D lesion segmentation and RECIST assessment in longitudinal studies.

\end{abstract}
\section{Introduction}
When reading computed tomography (CT) scans, an important step for radiologists is to measure the size of found lesions. Accurately measuring the lesions is crucial in precisely assessing the lesion growth rates across different time points. As such, it plays an important role in monitoring disease progression and making therapeutic plan. Currently, the most widely-used clinical guideline of lesion measurement is bidimensional RECIST (Response Evaluation Criteria In Solid Tumors) \cite{eisenhauer2009new} diameters. Manually labeling the RECIST diameters is tedious, time-consuming, and prone to be inconsistent among different observers \cite{yoon2016observer}. To overcome these issues, Tang \etal \cite{tang2018semi} first presented a semi-automatic RECIST diameter labeling approach. It has been demonstrated that segmentation can be considered as a more precise measurement than RECIST diameters to assess treatment response \cite{rothe2013size,bretschi2014assessing}. Therefore, many approaches \cite{cai2018accurate,tang2018ct,agarwal2020weakly1,agarwal2020weakly,tang2020one,tang2021weakly,tang2021lesion} have been proposed for lesion segmentation. Meanwhile, some great progresses \cite{tang2020one,tang2021lesion} have been made recently to further improve the performance of RECIST diameter prediction and incorporate lesion segmentation simultaneously.

The previous works \cite{tang2020one,tang2021lesion} tried to build powerful convolutional neural networks (CNN) through sophisticated designs to learn discriminative features. Although they achieved promising performances on lesion RECIST diameter prediction and segmentation (LRDPS), there are two obvious limitations. First, they failed to fully exploit the features' long-range dependencies due to the inherent locality of convolutional operations, which is crucial for pixel-wise prediction tasks, \eg lesion segmentation and heatmap prediction. Second, they only produced a single RECIST diameter output based on heatmap prediction. This would lead to unsatisfactory results when the lesions are difficult to identify.

On the other hand, transformer is designed to model sequence-to-sequence predictions in natural language processing tasks \cite{vaswani2017attention}, which is able to capture the long-range dependencies of the input sequences. Recently, it has been successfully applied to tasks of computer vision \cite{dosovitskiy2020image,carion2020end} and medical image segmentation \cite{cao2021swin,li2021medical,zhang2021transfuse}, which attracts tremendous attentions increasingly. 
Hence, to address the above limitations, we propose a transformer-based network (MeaFormer, \textbf{Mea}surement trans\textbf{Former}) for LRDPS in this paper, where three correlative and comprementary tasks are involved in including lesion segmentation, heatmap prediction, and keypoint regression.
Three main contributions are made in MeaFormer to push it into achieving accurate and robust results of LRDPS:
(1) transformers are employed to model the long-range dependencies of the high-resolution features extracted by a small CNN backbone. By incorporating rich contextual information, the enhanced high-resolution features facilitate accurate lesion segmentation and heatmap prediction results. (2) To the best of knowledge, it is the first time to use keypoint regression for RECIST diameter prediction and obtain RECIST diameters directly from segmentations. Thus, three types of RECIST diameter predictions are produced by MeaFormer in total instead of a single one. Fusing them helps to achieve a more robust outcome. (3) Two consistency losses are introduced to explicitly build relationships among these tasks. A joint optimization using them facilitates MeaFormer with an more optimal solution. Furthermore, extensive experiments on the large-scale DeepLesion dataset show that the proposed MeaFormer achieves the state-of-the-art performance of LRDPS.

\section{Method}
\begin{figure}[t]
  \centering
  \includegraphics[width=0.99\linewidth]{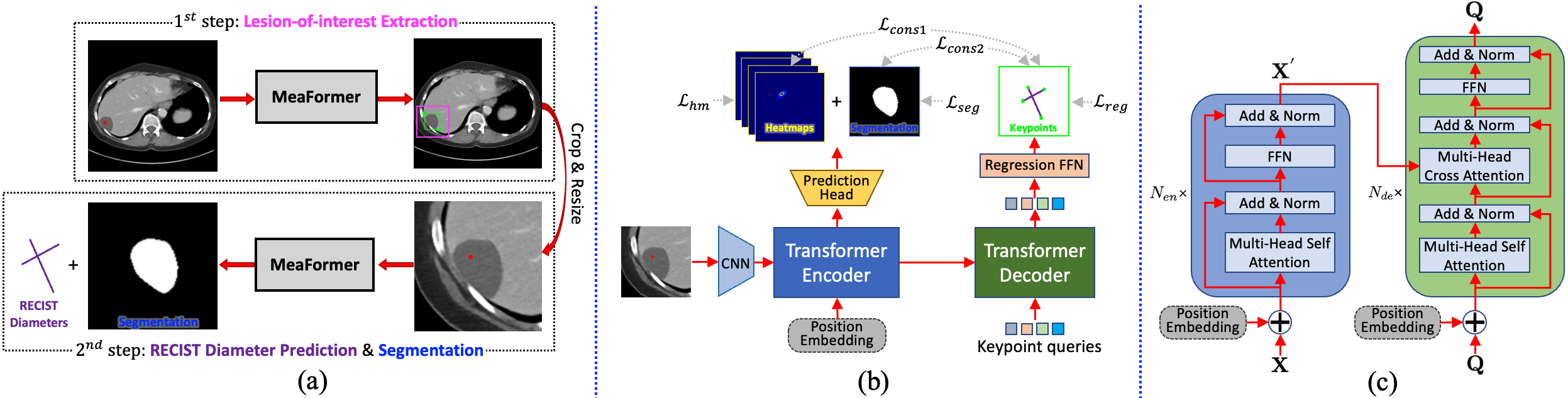}
  \caption{System overview. (a) The pipeline of our click-based automatic LRDPS system with separately trained MeaFormer models for step 1 and step 2, respectively. (b) The architecture of the proposed network MeaFormer used at step 2. (c) The detailed structures of MeaFormer's encoder and decoder. The MeaFormer used at step 1 has the same architecture except the following differences: (i) two queries for box prediction at step 1 rather than four at step 2 and (ii) no consistency losses (\ie $\mathcal{L}_{cons1}$ and $\mathcal{L}_{cons2}$) used for training at step 1. All figures are best viewed in color.}
  \label{fig:framework}
\end{figure}

As done in \cite{tang2020one,tang2021lesion}, we also develop a click-based automatic LRDPS system with two steps, whose pipeline is presented in Fig. \ref{fig:framework}(a). Given a CT image and a click guidance (the red spot) within the lesion, the top-left and bottom-right corners (the green circles) of a bounding box (the green dotted box) are predicted to indicate the lesion region at the $1^{st}$ step. A lesion-of-interest (LOI, the region in the magenta box) is extracted, whose center is the predicted box and whose width is two times the extent of the predicted box’s long side. As such, sufficient context information is preserved in a LOI. The LOI is then cropped and resized to $256 \times 256$. At the $2^{nd}$ step, the results of LRDPS are obtained based on the resized LOI. The proposed network MeaFormer is used at both steps.

\subsection{The Architecture of MeaFormer}
Fig. \ref{fig:framework}(b) shows the architecture of MeaFormer. We borrowed idea from DETR \cite{carion2020end} where transformers are applied on a CNN-based feature map instead of the image input. MeaFormer contains four main components, including a CNN backbone for feature extraction, an encoder-decoder transformer, a CNN-based prediction head for lesion segmentation and heatmap prediction, and a regression feed-forward network (FFN) for box/keypoint prediction at the $1^{st}$/$2^{nd}$ step. 

\noindent \textbf{Backbone.} Given an image $\mathbf{I}_{ct} \in \mathbb{R}^{H_0 \times W_0}$ of
 an original CT image at the $1^{st}$ step or a resized LOI CT image at the $2^{nd}$ step, we first generate a click image $\mathbf{I}_c$ and a distance transform image $\mathbf{I}_d$ based on the human click guidance following \cite{tang2020one,tang2021lesion}. A 3-channel image $\mathbf{I} \in \mathbb{R}^{3 \times H_0 \times W_0}$ is constructed by concatenating $\mathbf{I}_{ct}$, $\mathbf{I}_c$, and $\mathbf{I}_d$, which serves as an input of the backbone. To reduce the computation cost at the $1^{st}$ step, the original CT image is resized to $256 \times 256$. Thus, $H_0=W_0=256$ in this work. Here, we choose HRNet-W48 \cite{wang2020deep} as the backbone for feature extraction. As we know that the features' spatial and low-level information are crucial to getting accurate pixel-wise predictions, thus only the first three stages of HRNet-W48 is adopted to generate a low-level and high-resolution feature map $\mathbf{F} \in \mathbb{R}^{C \times H \times W}$, where $C = 48, H = \frac{H_0}{4}, W = \frac{W_0}{4}$.

\noindent \textbf{Encoder-Decoder Transformer.}
 We use a vanilla transformer architecture \cite{vaswani2017attention} for our transformer encoder and decoder design, whose detailed structures are shown in Fig. \ref{fig:framework}(c). Since the encoder takes as input a sequence, the feature map $\mathbf{F}$ is flattened and permuted to a sequence $\mathbf{X} \in \mathbb{R}^{(H \cdot W) \times C}$. In this work, the encoder has $N_{en} = 6$ layers. Each encoder layer contains a multi-head self-attention module and a feed-forward network (FFN). To make use of the spatial order of the sequence, the fixed positional encodings are added to the input sequence of each encoder layer. A new sequence $\mathbf{X}^{'} \in \mathbb{R}^{(H \cdot W) \times C}$ is produced after forwarding $N_{en}$ encoder layers. The decoder aims to transform a learnable query matrix $\mathbf{Q} \in \mathbb{R}^{N_{q} \times C}$ that contains $N_{q}$ query embeddings of size $C$ using multi-head self- and encoder-decoder attention mechanisms. It consists of $N_{de} = 6$ layers. Each layer takes as input $\mathbf{X}^{'}$ and $\mathbf{Q}$, and outputs an updated $\mathbf{Q}$. Similarly to the encoder, the positional encodings are added to the input of each decoder layer. After forwarding $N_{de}$ decoder layers, the final updated query matrix $\mathbf{Q}$ is token as input of the regression FFN. We set $N_{q}=2$ and $N_{q}=4$ at the $1^{st}$ and $2^{nd}$ step, respectively. That is because two corners/four keypoints are required for box/keypoint prediction.

\noindent \textbf{Prediction Head.}
A CNN-based prediction head is attached to the transformer encoder for lesion segmentation and heatmap prediction. Before feeding the sequence $\mathbf{X}^{'} \in \mathbb{R}^{(H \cdot W) \times C}$ into the prediction head, we first permute and reshape $\mathbf{X}^{'}$ to a feature map $\mathbf{F}^{'} \in \mathbb{R}^{C \times H \times W}$. The prediction head consists of sequential layers: $conv(kn=32,ks=3,st=1)$, $deconv(32,4,2)$, $conv(32,3,1)$, $deconv(32,4,2)$, and $conv(5,1,1)$, where $conv(*)$, $deconv(*)$, $kn$, $ks$, and $st$ represent convolutional layer, deconvolutional layer, kernel number, kernel size, and stride, respectively. Each layer is followed by a batch normalization and a ReLU except the last one. Therefore, the output of the prediction head $\mathbf{O} \in \mathbb{R}^{5 \times H_0 \times W_0}$ has the same resolution to the input image. One channel of $\mathbf{O}$ is the lesion segmentation result $\mathbf{S}$ and the other four channels are the predicted heatmaps $\mathbf{M}$.

\noindent \textbf{Regression FFN.}
To get the box/keypoint prediction results, a regression FFN is attached to the transformer decoder. It consists of a 2-layer perceptron with ReLU activation function and hidden dimension $d=96$, and a linear projection layer with two output nodes for predicting the corner/keypoint coordinate $(x,y)$ using a softmax function. As done in \cite{carion2020end}, we also add the regression FFN after each transformer decoder layer to introduce auxiliary losses for model optimization. All regression FFNs share their parameters. 

\subsection{Model Optimization}
The porposed MeaFormer has three task-related losses to optimize. The lesion segmentation loss $\mathcal{L}_{seg}$ is the summation of a binary cross entropy loss $\mathcal{L}_{bce}$ and an IoU loss $\mathcal{L}_{iou}$ \cite{rahman2016optimizing}, \ie $\mathcal{L}_{seg} = \mathcal{L}_{bce} + \mathcal{L}_{iou}$. 
The same procedure as \cite{tang2021weakly} is conducted to construct initial and updated lesion pseudo masks as supervision for $\mathcal{L}_{seg}$ optimization. 
For heatmap prediction, the objective function $\mathcal{L}_{hm}$ is a mean squared error loss $\mathcal{L}_{mse}$ that measures the errors between the predicted heatmaps and the ground truth heatmaps, which are four 2D Gaussian maps (with a standard deviation of 5 pixels) centered on the endpoints of RECIST annotations. For keypoint regression, the objective function $\mathcal{L}_{reg}$ is a L1 loss $\mathcal{L}_1$ that is adopted to calculate the errors between the predicted keypoint coordinates produced by $N_{de}$ regression FFNs and the ground truth coordinates of the RECIST diameters' endpoints.

Besides these individual losses, this work also introduces two consistency losses to explicitly build two relationships among different tasks. One relationship is built between the tasks of keypoint regression and heatmap prediction. Both of them are conducted for RECIST diameter prediction, thus their outputs should be as consistent as possible. Let $(x_i,y_i)$ denote the $i^{th}$ keypoint's predicted coordinate produced by the keypoint regression task, so $\mathbf{M}_{i,y_i,x_i}$ means the value of the $i^{th}$ predicted heatmap at location $(x_i,y_i)$. To make the outputs of keypoint regression and heatmap prediction as consistent as possible, $(x_i,y_i)$ should be as close as possible to the maximum location of the $i^{th}$ predicted heatmap, meaning that the error between $\mathbf{M}_{i,y_i,x_i}$ and 1 should be as small as possible. The other relationship is built between the tasks of keypoint regression and lesion segmentation. In an ideal situation, the predicted keypoints should be located at the boundaries of the lesion segmentation. We first binarize the segmentation result by $\mathbf{S} \geq 0.5$ , and then perform a distance transformation on the binary mask to get a distance map $\mathbf{D}$. Therefore, to make the results of keypoint regression and lesion segmentation as consistent as possible, the error between $\mathbf{D}_{y_i,x_i}$ and 0 should be as small as possible. L1 losses are used to compute the aforementioned two errors, which are denoted as $\mathcal{L}_{cons1}$ and $\mathcal{L}_{cons2}$, respectively.
The final objective function for model optimization is defined as
\begin{equation}
    \mathcal{L} = \lambda_1 \mathcal{L}_{seg} + \lambda_2 \mathcal{L}_{hm} + \lambda_3 \mathcal{L}_{reg} + \lambda_4 \mathcal{L}_{cons1} + \lambda_5 \mathcal{L}_{cons2},
\end{equation}
where we set $\lambda_1=1$, $\lambda_2=10$, $\lambda_3=1$, $\lambda_4=0.01$, and $\lambda_5=0.01$ to balance the magnitude of different losses in this work.

\noindent \textbf{Implementation Details.}
MeaFormer is implemented in PyTorch 1.6 \cite{paszke2019pytorch} and the CNN backbone is initialized with \mbox{ImaeNet} \cite{DengDSLL009} pre-trained weights. Adam optimizer \cite{kingma2014adam} with an initial learning rate of 0.001 and default settings (\eg betas=(0.9, 0.999)) is used to train the model for 200 epochs reduced by 0.1 at epoch 100 and 150. The training batch size is 16. The following operations are randomly conducted for data augmentation: scaling, cropping, rotating, brightness and contrast adjusting, and Gaussian blurring. The MeaFormer models used at two steps are trained separately. Following \cite{tang2021weakly}, the one at the $2^{nd}$ step is trained three rounds with the iteratively updated lesion pseudo masks as supervisions for weakly-supervised lesion segmentation learning.

\setcounter{footnote}{0}

\section{Experiments}
\textbf{Dataset and Evaluation Metrics.}
A large-scale DeepLesion dataset \cite{yan2018deeplesion}
\footnote{https://nihcc.app.box.com/v/DeepLesion} 
is used to train and test the proposed MeaFormer. It collects $32,735$ lesions from $10,594$ studies of $4,427$ patients. There are a variety of lesion types in this dataset, such as lung nodules, liver tumors, enlarged lymph nodes, and so on. All lesion are annotated by bidimensional RECIST diameters telling their sizes and locations. As done in \cite{tang2020one,tang2021lesion}, $1,000$ lesion images from 500 patients are manually segmented as a test set. The rest patient data are used for training (80\%) and validation (20\%). The same evaluation metrics to \cite{tang2020one,tang2021lesion} are adopted to calculate the quantitative results, including the pixel-wise dice coefficient (Dice) for lesion segmentation and the differences between the diameter lengths (mm) of the predictions and manual annotations for RECIST diameter prediction.

\begin{figure}[t]
    \centering
    \includegraphics[width=0.99\textwidth]{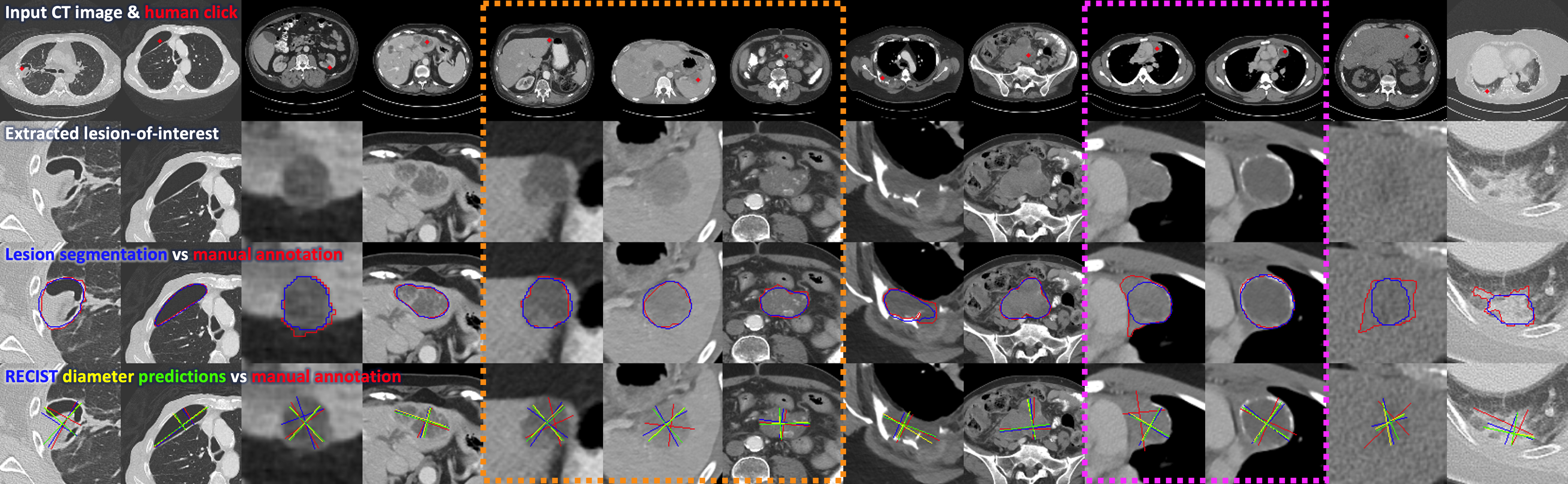}
    \caption{Qualitative results of our system. Each column gives an example, where the content of each row has been described with texts. The predicted RECIST diameters shown in the fourth row with different colors are obtained from three different tasks, including \textcolor{blue}{segmentation}, \textcolor{yellow}{heatmap prediction}, and \textcolor{green}{keypoint regression}. The orange dotted box shows three examples with imperfect manual diameter annotations. The magenta dotted box shows a lesion pair collected from the same patient at different time points, where the manual annotations of the left lesion are incorrect.} 
    \label{fig:result}
\end{figure}

\begin{table}[t!]
	\begin{center}
		\caption{Results of different methods for LRDPS on the DeepLesion test set in terms of model size (\ie the parameter number, the unit is million), Dice ($\%$), and long/short diameter length error (mm). The mean and standard deviation are reported. \textcolor{red}{Red} and \textcolor{blue}{blue} texts indicate rank 1 and 2, respectively.}
		\label{tab:result}
		{
			\scriptsize
			
			\begin{tabu}{| X[2.1c] | X[0.7c] | X[1.2c] | X[c] | X[c] | X[c] | X[c] | X[c] | X[c] | X[c] | X[c] |}
				\hline
				 \multirow{2}{*}{Method} & \multirow{2}{*}{Size} & \multirow{2}{*}{Dice} & \multicolumn{2}{c|}{Heatmaps}  & \multicolumn{2}{c|}{Segmentation} & \multicolumn{2}{c|}{Regression}  & \multicolumn{2}{c|}{Fusion} \\ \cline{4-11}
                 & & & Long & Short & Long & Short & Long & Short & Long & Short \\ \hline
				Cai \etal \cite{cai2018accurate}  & - &  90.6$\pm$8.9  & -  & -   & - &  - & - &  - & - &  -\\ 
				AHRNet \cite{tang2021weakly}                           & - &  \textcolor{blue}{92.6$\pm$4.3}  &  - & -   & - &  - & - &  - & - &  -\\  \hline
				Tang \etal \cite{tang2018semi}  & - &  - & 1.9$\pm$2.2 &  1.6$\pm$1.9  & - &  - & - &  - & - &  -\\ \hline
				SEENet \cite{tang2020one}  & - & 91.2$\pm$3.9 &  1.7$\pm$2.0 & 1.6$\pm$1.8  & - &  - & - &  - & - &  -\\ 
				nnUNet \cite{isensee2018nnu}  & - &  90.7$\pm$5.0 &  2.1$\pm$2.0 & 1.8$\pm$1.7  & - &  - & - &  - & - &  -\\ 
				PDNet \cite{tang2021lesion}  & - &  92.4$\pm$4.5 &  \textcolor{blue}{1.7$\pm$1.5} & \textcolor{blue}{1.5$\pm$1.4}  & - &  - & - &  - & - &  -\\ \hline
				Swin-Unet \cite{cao2021swin} & 84.0 &  91.0$\pm$5.1 &  1.9$\pm$1.9 & 2.0$\pm$3.7  & 2.0$\pm$1.8 &  1.8$\pm$2.2 & - &  - & 1.9$\pm$1.8 &  1.7$\pm$2.5\\ 
				SegTran \cite{li2021medical} & 134.8 &  91.9$\pm$5.2 &  1.8$\pm$1.5 & 1.9$\pm$3.1  & 1.8$\pm$1.7 &  1.7$\pm$2.1 & - &  - & 1.7$\pm$1.6 &  1.7$\pm$2.2\\ 
				TransFuse \cite{zhang2021transfuse} & \textcolor{blue}{26.1} &  92.0$\pm$4.9 &  1.8$\pm$1.4 & 1.8$\pm$2.0  & \textcolor{blue}{1.7$\pm$1.6} &  \textcolor{blue}{1.5$\pm$2.1} & - &  - & \textcolor{blue}{1.7$\pm$1.5} &  \textcolor{blue}{1.6$\pm$1.8}\\ \hline
				Ours  & \textcolor{red}{18.3} &  \textcolor{red}{92.7$\pm$4.3} &  \textcolor{red}{1.6$\pm$1.4}  &  \textcolor{red}{1.4$\pm$1.6}  & \textcolor{red}{1.6$\pm$1.3} &  \textcolor{red}{1.5$\pm$1.5} & \textcolor{red}{1.9$\pm$1.5}  &  \textcolor{red}{1.7$\pm$1.9} & \textcolor{red}{1.6$\pm$1.3} &  \textcolor{red}{1.4$\pm$1.5}\\ 
				\hline
			\end{tabu}
		}
	\end{center}
\end{table}

\noindent \textbf{Experimental Results.}
Fig. \ref{fig:result} displays several qualitative results produced by our system. We observe that (1) the lesions are well located at the center of the extracted LOIs, meaning that the proposed network can accurately predict the corners of the lesion bounding boxes. A prediction is considered as correct when the IoU of the predicted box and the ground truth box is larger than 0.5. On the 1000 test lesions, we achieve an accuracy of $99.1\%$ surpassing the previous work \cite{tang2020one} by about $2\%$. (2) Our automatic results are close to the manual annotations and the diameter predictions produced by different tasks are nearly consistent, which visually demonstrate the effectiveness of our system and the contributions of the proposed consistency losses. (3) For some cases (\eg the examples in the orange dotted box), even the manual diameter annotations are not very precise, our automatic predictions can mitigate this issue. (4) For some cases (the left example in the magenta dotted box), the lesions' boundaries are unclear for some areas, resulting in incorrect manual annotations, but the automatic results can correct them. (5) Although our system works well on most of cases, it still faces some difficulties when the lesions' boundaries/shapes are heavily vague/irregular (the examples in the last two columns).

The quantitative comparisons of our system with previous approaches on these tasks \cite{cai2018accurate,tang2018semi,isensee2018nnu,tang2020one,tang2021weakly,tang2021lesion} and three transformer-based medical image segmentation approaches \cite{cao2021swin,li2021medical,zhang2021transfuse} are presented in Table \ref{tab:result}. The listed results of \cite{cai2018accurate,tang2018semi,tang2020one,tang2021weakly,tang2021lesion} are copied from their related papers. The results of nnUnet \cite{isensee2018nnu} are copied from \cite{tang2021lesion}. For \cite{cao2021swin,li2021medical,zhang2021transfuse}, we train them for heatmap prediction and lesion segmentation at the $2^{nd}$ step using the same data as ours for fair comparisons. We observe that (1) Our MeaFormer has the smallest model size but achieves the best performance on all tasks, demonstrating its effectiveness for these tasks. (2) Better segmentation results produce better RECIST diameter predictions from segmentation. Sometimes, it even outperforms the heatmap-based prediction. This demonstrates that segmentation is a reliable way to predict RECIST diameters. Specifically, the RECIST diameters are obtained by first finding the longest diameter (long axis) from a lesion segmentation, and then calculating its longest perpendicular diameter (short axis). (3) More reliable RECIST diameter lengths can be obtained after fusing the predictions from different tasks, which mimics the behavior of radiologists making a consistent decision together when they have different observations. In this work, we find that the following fusion strategy achieves the best performance: (i) use prediction from segmentation as the reference, (ii) select a prediction from another task that has the smallest length difference compared to the reference, (iii) average their lengths to estimate the lesion size.
\begin{figure}[t]
    \centering
    \includegraphics[width=0.99\textwidth]{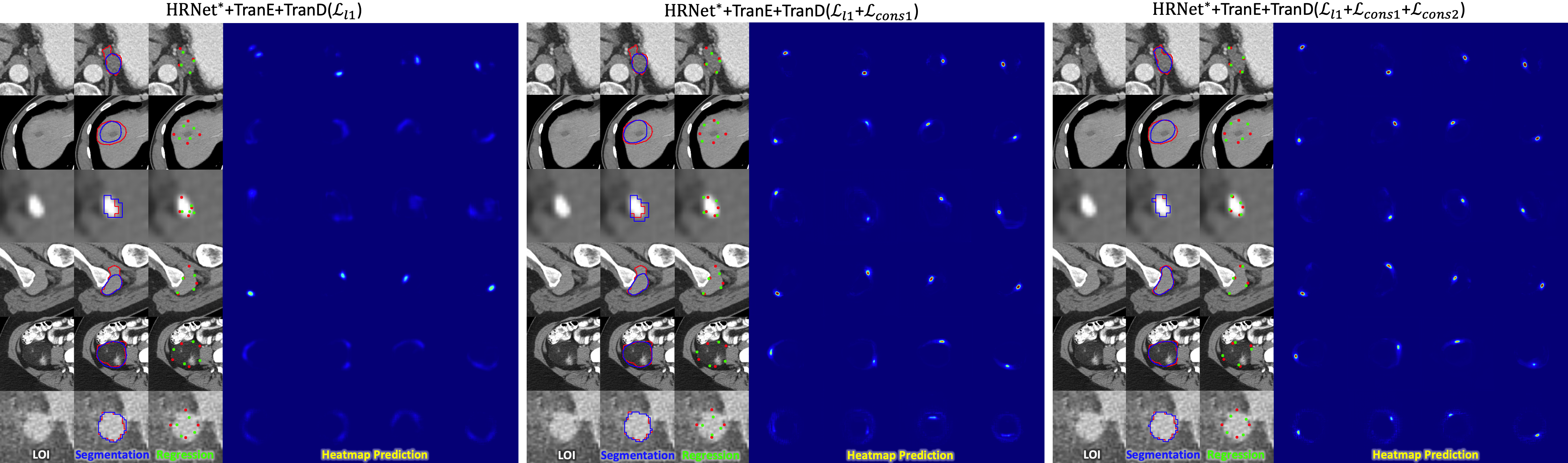}
    \caption{Visual comparisons of the results produced by the proposed network that is trained with or without the consistency losses. The red curves and spots are the \textcolor{red}{manually annotated lesion masks and RECIST diameters' endpoints}. The blue curves are the \textcolor{blue}{automatic lesion segmentation results}. The green spots are the \textcolor{green}{predicted RECIST diameters' endpoints by keypoint regression}. } 
    \label{fig:ablation}
\end{figure}

\begin{table}[!t]
    \begin{center}
    {
		\caption{Results of different settings of our network on the DeepLesion test set.}
		\label{tab:ablation}
		\scriptsize
        \begin{tabu} to 0.99\textwidth {| X[5.2c] | X[0.6c] | X[c] | X[0.9c] | X[0.9c] | X[0.9c] | X[0.9c] |}
              \hline
              \multirow{2}{*}{Method} & \multirow{2}{*}{Size} & \multirow{2}{*}{Dice} & \multicolumn{2}{c|}{Heatmaps}  & \multicolumn{2}{c|}{Regression}  \\ \cline{4-7}
                 &  & & Long & Short & Long & Short \\ \hline
              HRNet$_{regression}$   & 39.2 &  - & -  &  - & 3.0$\pm$2.3  &  2.1$\pm$2.2 \\ 
              HRNet$_{heatmap}$  & 70.4 &  91.4$\pm$5.7& 1.9$\pm$1.7  &  1.9$\pm$2.4 & -  &  - \\ 
        	  HRNet$^*$+TranE  & \textcolor{red}{17.6} &  92.1$\pm$4.9 & 1.7$\pm$1.6  &  1.7$\pm$2.1 & -  &  - \\ 
        	  HRNet$^*$+TranE+TranD($\mathcal{L}_1$)  & \textcolor{blue}{18.3} &  \textcolor{blue}{92.5$\pm$4.4} & 1.7$\pm$1.7  &  1.6$\pm$1.7 & 2.3$\pm$1.9  &  2.0$\pm$2.2 \\
        	  HRNet$^*$+TranE+TranD($\mathcal{L}_1$+$\mathcal{L}_{cons1}$)  & \textcolor{blue}{18.3} &  92.5$\pm$4.5 & \textcolor{blue}{1.7$\pm$1.5}  &  \textcolor{blue}{1.5$\pm$1.7} & \textcolor{blue}{1.9$\pm$1.6}  &  \textcolor{blue}{1.8$\pm$2.0} \\
        	  HRNet$^*$+TranE+TranD($\mathcal{L}_1$+$\mathcal{L}_{cons1}$+$\mathcal{L}_{cons2}$)  & \textcolor{blue}{18.3} &  \textcolor{red}{92.7$\pm$4.3} & \textcolor{red}{1.6$\pm$1.4}  &  \textcolor{red}{1.4$\pm$1.6} & \textcolor{red}{1.9$\pm$1.5}  &  \textcolor{red}{1.7$\pm$1.9} \\
              \hline
        \end{tabu}
    }
    \end{center}
\end{table}%

\noindent \textbf{Ablation Studies.}
The proposed network is built based on HRNet by gradually integrating a transformer encdoer (TranE) and a transformer decoder (TranD). Besides, two new consistency losses ($\mathcal{L}_{cons1}$ and $\mathcal{L}_{cons2}$) are introduced for model optimization. Although only the first three stages of HRNet (HRNet$^*$) are used in this work, we also adopt the full HRNet for keypoint regression (HRNet$_{regression}$) by adding a global average pooling layer and a fully connected layer, and lesion segmentation and heatmap prediction (HRNet$_{heatmap}$) by adding a UNet-like decoder and skip connections. Table \ref{tab:ablation} lists the results of different settings. We observe that (1) the performance of all tasks are consistently improved after gradually introducing different components, demonstrating the effectiveness and contributions of each component for these tasks. (2) After simply adding a transformer encoder (HRNet$^*$+TranE), the performance gets a big improvement compared to HRNet$_{heatmap}$, meaning that the encoder can well capture the long-range dependencies of low-level features to enhance their representation ability. (3) After introducing the consistency losses, all tasks' results are improved, especially for keypoint regression when adding $\mathcal{L}_{cons1}$ and lesion segmentation when adding $\mathcal{L}_{cons2}$. Fig. \ref{fig:ablation} provides visual evidences to show the effectiveness of the proposed consistency losses, where the results of lesion segmentation and keypoint regression become closer to manual annotations and the predicted heatmaps become more concentrated and reliable after adding them.

\noindent \textbf{Downstream Tasks.}
Two straightforward downstream clinic-relevant tasks can be conducted based on our results, which are 3D lesion segmentation and RECIST assessment in longitudinal studies. For 3D lesion segmentation, 200 lesions selected from above manually segmented 1,000 lesions are fully annotated with 3D masks following \cite{cai2018accurate}. We run our system on these lesions slice-by-slice and stack their 2D segmentation results to form their 3D segmentations. This work achieves a mean Dice score of $85.6\%$ remarkably surpassing the previous work \cite{cai2018accurate} by $9.2\%$. For RECIST assessment in longitudinal studies, 480 lesion pairs from the test set of the DLS dataset \cite{cai2021deep}
\footnote{https://github.com/JimmyCai91/DLT} 
are used for evaluation, which are excluded from our training data. The CT scans of each pair are collected at different time points from the same patient. For each lesion pair, we first get the lesions' RECIST diameters using our system. Then we compute its tumor response class, as complete response, partial response, progressive disease, or stable disease, of the manual RECIST and our automated RECIST assessments, based on RECIST version 1.1 \cite{eisenhauer2009new}. On the 480 test lesion pairs, we achieve an accuracy of $91.7\%$ for the tumor response classification, suggesting that our system can help clinicians to annotate sequential RECIST measurements with minimal human effort. 

\section{Conclusions}
In this work, we propose a transformer-based network for lesion RECIST diameter prediction and segmentation. It leverages the long-range dependencies learning ability of transformers to enhance features. Besides, three tasks are performed and jointly trained, and their relationships are explicitly built by introducing two new consistency losses. All of these facilitate our system to produce accurate and robust results. Hence, it can serve as a useful tool for clinicians to precisely measure tumor sizes with minimal effort and may provide high positive clinical values on oncologic imaging analysis workflows.

\bibliographystyle{splncs04}
\bibliography{egbib}

\end{document}